\documentclass[longauth]{aa}
\usepackage{graphicx}
\usepackage{txfonts}
\usepackage{xcolor}
\usepackage{hyperref}
\hypersetup{colorlinks, linkcolor=blue, citecolor=blue, urlcolor=blue}
\newcommand{\te}{t_{\rm E}}
\newcommand{\thetae}{\theta_{\rm E}}

\newcommand{\dl}{D_{\rm L}}
\newcommand{\ds}{D_{\rm S}}

\definecolor{brown}{rgb}{0.59, 0.29, 0.0}
\definecolor{darkgreen}{rgb}{0.0, 0.42, 0.24}
\definecolor{darkblue}{rgb}{0.01, 0.31, 0.59}
\definecolor{darkblue}{rgb}{0.0, 0.25, 0.42}
\definecolor{blue}{rgb}{0.0,0.0,1.0}
\definecolor{green}{rgb}{0.0,1.0,0.0}

\begin{document}

\title{
MOA-2022-BLG-091Lb and KMT-2024-BLG-1209Lb:
Microlensing planets detected through weak caustic-crossing signals
}
\titlerunning{MOA-2022-BLG-091Lb and KMT-2024-BLG-1209Lb}

\author{
     Cheongho~Han\inst{\ref{cbnu}}
\and Chung-Uk~Lee\inst{\ref{kasi}\thanks{\tt leecu@kasi.re.kr}} 
\and Andrzej~Udalski\inst{\ref{warsaw}} 
\and Ian~A.~Bond\inst{\ref{massey}}
\and Hongjing~Yang\inst{\ref{tsinghua}}     
\\
(Leading authors)
\\
     Michael~D.~Albrow\inst{\ref{canterbury}}   
\and Sun-Ju~Chung\inst{\ref{kasi}}      
\and Andrew~Gould\inst{\ref{osu}}      
\and Youn~Kil~Jung\inst{\ref{kasi},\ref{ust}} 
\and Kyu-Ha~Hwang\inst{\ref{kasi}} 
\and Yoon-Hyun~Ryu\inst{\ref{kasi}} 
\and Yossi~Shvartzvald\inst{\ref{weizmann}}   
\and In-Gu~Shin\inst{\ref{cfa}} 
\and Jennifer~C.~Yee\inst{\ref{cfa}}   
\and Weicheng~Zang\inst{\ref{cfa},\ref{tsinghua}}     
\and Tanagodchaporn Inyanya\inst{\ref{kasi},\ref{ust}}
\and Sang-Mok~Cha\inst{\ref{kasi},\ref{kyunghee}} 
\and Doeon~Kim\inst{\ref{cbnu}}
\and Dong-Jin~Kim\inst{\ref{kasi}} 
\and Seung-Lee~Kim\inst{\ref{kasi}} 
\and Dong-Joo~Lee\inst{\ref{kasi}} 
\and Yongseok~Lee\inst{\ref{kasi},\ref{kyunghee}} 
\and Byeong-Gon~Park\inst{\ref{kasi}} 
\and Richard~W.~Pogge\inst{\ref{osu}}
\\
(The KMTNet Collaboration)
\\
     Przemek~Mr{\'o}z\inst{\ref{warsaw}} 
\and Micha{\l}~K.~Szyma{\'n}ski\inst{\ref{warsaw}}
\and Jan~Skowron\inst{\ref{warsaw}}
\and Rados{\l}aw~Poleski\inst{\ref{warsaw}} 
\and Igor~Soszy{\'n}ski\inst{\ref{warsaw}}
\and Pawe{\l}~Pietrukowicz\inst{\ref{warsaw}}
\and Szymon~Koz{\l}owski\inst{\ref{warsaw}} 
\and Krzysztof~A.~Rybicki\inst{\ref{warsaw},\ref{weizmann}}
\and Patryk~Iwanek\inst{\ref{warsaw}}
\and Krzysztof~Ulaczyk\inst{\ref{warwick}}
\and Marcin~Wrona\inst{\ref{warsaw},\ref{villanova}}
\and Mariusz~Gromadzki\inst{\ref{warsaw}}          
\and Mateusz~J.~Mr{\'o}z\inst{\ref{warsaw}} 
\and Micha{\l} Jaroszy{\'n}ski\inst{\ref{warsaw}}
\and Marcin Kiraga\inst{\ref{warsaw}}
\\
(The OGLE Collaboration)
\\
     Fumio~Abe\inst{\ref{nagoya}}
\and Ken~Bando\inst{\ref{osaka}}
\and David~P.~Bennett\inst{\ref{goddard},\ref{maryland}}
\and Aparna~Bhattacharya\inst{\ref{goddard},\ref{maryland}}
\and Akihiko~Fukui\inst{\ref{tokyo-earth},}\inst{\ref{spain}}
\and Ryusei~Hamada\inst{\ref{osaka}}
\and Shunya~Hamada\inst{\ref{osaka}}
\and Naoto Hamasaki\inst{\ref{osaka}}
\and Yuki~Hirao\inst{\ref{tokyo-astro}}
\and Stela~Ishitani Silva\inst{\ref{goddard},\ref{oak-ridge}}  
\and Naoki~Koshimoto\inst{\ref{osaka}}
\and Yutaka~Matsubara\inst{\ref{nagoya}}
\and Shota~Miyazaki\inst{\ref{jaxa}}
\and Yasushi~Muraki\inst{\ref{nagoya}}
\and Tutumi Nagai\inst{\ref{osaka}}
\and Kansuke Nunota\inst{\ref{osaka}}
\and Greg~Olmschenk\inst{\ref{goddard}}
\and Cl{\'e}ment~Ranc\inst{\ref{sorbonne}}
\and Nicholas~J.~Rattenbury\inst{\ref{auckland}}
\and Yuki~Satoh\inst{\ref{osaka}}
\and Takahiro~Sumi\inst{\ref{osaka}}
\and Daisuke~Suzuki\inst{\ref{osaka}}
\and Sean K. Terry\inst{\ref{goddard}, \ref{maryland}}
\and Paul~J.~Tristram\inst{\ref{john}}
\and Aikaterini~Vandorou\inst{\ref{goddard},\ref{maryland}}
\and Hibiki~Yama\inst{\ref{osaka}}
\\
(The MOA Collaboration)
\\
     Yunyi~Tang\inst{\ref{tsinghua}}
\and Shude~Mao\inst{\ref{tsinghua}}
\and Dan~Maoz\inst{\ref{telaviv}}
\and Wei~Zhu\inst{\ref{tsinghua}}
\\
(The LCO Collaboration)
}

\institute{
      Department of Physics, Chungbuk National University, Cheongju 28644, Republic of Korea                                                          \label{cbnu}     
\and  Korea Astronomy and Space Science Institute, Daejon 34055, Republic of Korea                                                                    \label{kasi}   
\and  Astronomical Observatory, University of Warsaw, Al.~Ujazdowskie 4, 00-478 Warszawa, Poland                                                      \label{warsaw}   
\and  Institute of Natural and Mathematical Science, Massey University, Auckland 0745, New Zealand                                                    \label{massey}    
\and  Department of Astronomy and Tsinghua Centre for Astrophysics, Tsinghua University, Beijing 100084, China                                        \label{tsinghua} 
\and  University of Canterbury, Department of Physics and Astronomy, Private Bag 4800, Christchurch 8020, New Zealand                                 \label{canterbury}  
\and  Department of Astronomy, Ohio State University, 140 West 18th Ave., Columbus, OH 43210, USA                                                     \label{osu} 
\and  University of Science and Technology, Daejeon 34113, Republic of Korea                                                                          \label{ust}
\and  Department of Particle Physics and Astrophysics, Weizmann Institute of Science, Rehovot 76100, Israel                                           \label{weizmann}   
\and  Center for Astrophysics $|$ Harvard \& Smithsonian 60 Garden St., Cambridge, MA 02138, USA                                                      \label{cfa}  
\and  School of Space Research, Kyung Hee University, Yongin, Kyeonggi 17104, Republic of Korea                                                       \label{kyunghee}     
\and  Department of Physics, University of Warwick, Gibbet Hill Road, Coventry, CV4 7AL, UK                                                           \label{warwick}
\and  Villanova University, Department of Astrophysics and Planetary Sciences, 800 Lancaster Ave., Villanova, PA 19085, USA                           \label{villanova} 
\and  Institute for Space-Earth Environmental Research, Nagoya University, Nagoya 464-8601, Japan                                                     \label{nagoya}     
\and  Department of Earth and Space Science, Graduate School of Science, Osaka University, Toyonaka, Osaka 560-0043, Japan                            \label{osaka}  
\and  Code 667, NASA Goddard Space Flight Center, Greenbelt, MD 20771, USA                                                                            \label{goddard}  
\and  Department of Astronomy, University of Maryland, College Park, MD 20742, USA                                                                    \label{maryland}      
\and  Department of Earth and Planetary Science, Graduate School of Science, The University of Tokyo, 7-3-1 Hongo, Bunkyo-ku, Tokyo 113-0033, Japan   \label{tokyo-earth}    
\and  Instituto de Astrof{\'i}sica de Canarias, V{\'i}a L{\'a}ctea s/n, E-38205 La Laguna, Tenerife, Spain                                            \label{spain} 
\and  Institute of Astronomy, Graduate School of Science, The University of Tokyo, 2-21-1 Osawa, Mitaka, Tokyo 181-0015, Japan                        \label{tokyo-astro} 
\and  Oak Ridge Associated Universities, Oak Ridge, TN 37830, USA                                                                                     \label{oak-ridge} 
\and  Institute of Space and Astronautical Science, Japan Aerospace Exploration Agency, 3-1-1 Yoshinodai, Chuo, Sagamihara, Kanagawa 252-5210, Japan  \label{jaxa}  
\and  Sorbonne Universit\'e, CNRS, UMR 7095, Institut d'Astrophysique de Paris, 98 bis bd Arago, 75014 Paris, France                                  \label{sorbonne} 
\and  Department of Physics, University of Auckland, Private Bag 92019, Auckland, New Zealand                                                         \label{auckland}  
\and  University of Canterbury Mt.~John Observatory, P.O. Box 56, Lake Tekapo 8770, New Zealand                                                       \label{john}  
\and  School of Physics and Astronomy, Tel-Aviv University, Tel-Aviv 6997801, Israel                                                                  \label{telaviv}
}                                                                                                                                                       
\date{Received ; accepted}

% \abstract{}{}{}{}{} 
% 5 {} token are mandatory
\abstract
{}
{
The light curves of the microlensing  events MOA-2022-BLG-091 and KMT-2024-BLG-1209 exhibit 
anomalies with very similar features. These anomalies appear near the peaks of the light curves, 
where the magnifications are moderately high, and are distinguished by weak caustic-crossing 
features with minimal distortion while the source remains inside the caustic.  To achieve a 
deeper understanding of these anomalies, we conducted a comprehensive analysis of the lensing 
events.
}
{
We carried out binary-lens modeling with a thorough exploration of the parameter space.
This analysis revealed that the anomalies in both events are of planetary origin, although 
their exact interpretation is complicated by different types of degeneracy. In the case of 
MOA-2022-BLG-091, the main difficulty in the interpretation of the anomaly arises from a 
newly identified degeneracy related to the uncertain angle at which the source trajectory 
intersects the planet-host axis. For KMT-2024-BLG-1209, the interpretation is affected by 
the previously known inner-outer degeneracy, which leads to ambiguity between solutions 
in which the source passes through either the inner or outer caustic region relative to 
the planet host.
\color{black}
}
{
Bayesian analysis indicates that the planets in both lens systems are giant planets with 
masses about 2 to 4 times that of Jupiter, orbiting early K-type main-sequence stars. Both 
systems are likely located in the Galactic disk at a distance of around 4 kiloparsecs.
The degeneracy in KMT-2024-BLG-1209 is challenging to resolve because it stems from intrinsic 
similarities in the caustic structures of the degenerate solutions. In contrast, the degeneracy 
in MOA-2022-BLG-091, which occurs by chance rather than from inherent characteristics, is 
expected to be resolved by the future space based Roman RGES microlensing survey.
}
{}

\keywords{planets and satellites: detection -- gravitational lensing: micro}

\maketitle

\section{Introduction} \label{sec:one}

Gravitational microlensing is an important technique for detecting extrasolar planets, 
offering unique scientific value by enabling the discovery of planetary types that are 
difficult to detect using other methods \citep{Mao1991, Gould1992b}. Notably, microlensing 
is particularly sensitive to wide-orbit planets, which orbit their host stars at significant 
distances, such as OGLE-2016-BLG-1227L \citep{Han2020}, OGLE-2012-BLG-0838L \citep{Poleski2020}, 
and OGLE-2017-BLG-0448L \citep{Zhai2024}. It is also adept at detecting free-floating planets 
that are not gravitationally bound to any star, including OGLE-2012-BLG-1323L and 
OGLE-2017-BLG-0560L \citep{Mroz2019}, OGLE-2019-BLG-0551L \citep{Mroz2020}, OGLE-2016-BLG-1540L 
\citep{Mroz2018}, KMT-2019-BLG-2073L \citep{Kim2021}, and KMT-2023-BLG-2669L \citep{Jung2024}. 
By revealing these elusive objects, microlensing plays a pivotal role in advancing our 
understanding of planetary demographics and complements the discoveries made by other detection 
techniques.

Currently, planetary microlensing surveys are being conducted by three major groups: the Optical
Gravitational Lensing Experiment \citep[OGLE;][]{Udalski2015}, the Microlensing Observations in
Astrophysics \citep[MOA;][]{Bond2001, Sumi2003}, and the Korea Microlensing Telescope Network 
\citep[KMTNet;][]{Kim2016}. These surveys utilize a network of 1-meter-class ground-based 
telescopes equipped with wide-field cameras, strategically distributed across multiple locations 
in the Southern Hemisphere. By concentrating on the densely populated stellar regions near the
Galactic center, they detect microlensing events and identify planetary signals embedded within 
the light curves of these events. Collectively, these efforts yield approximately 30 planetary 
discoveries annually \citep{Gould2022}, making microlensing the third most productive planet 
detection method, after the transit and radial velocity techniques.

Gravitational microlensing surveys are planned to undergo further expansion. One is the
PRime-focus Infrared Microlensing Experiment (PRIME), a recently completed project currently
undergoing test observations. This survey aims to conduct a dedicated near-infrared microlensing
survey using a 1.8-meter telescope with a wide field of view of 1.45 square degrees, located at the
South African Astronomical Observatory. Its primary goal is to observe the central bulge region of
the Milky Way, where high extinction makes optical observations challenging. The PRIME survey is
expected to detect 42--52 planets annually, including 1--2 planets with masses below that of Earth,
offering a significant contribution to the study of planetary demographics in the high-extinction
regions of our galaxy \citep{Kondo2023}.

The Nancy Grace Roman Space Telescope (Roman) is set to conduct the Galactic Exoplanet Survey
(RGES) from space, employing the microlensing method to discover planets. The planned survey
area spans approximately 2 square degrees near the Galactic bulge, utilizing the telescope's 
wide 1-2 $\mu$m W146 filter with a 15-minute observational cadence. Over the course of six 72-day
observing seasons, the RGES is expected to identify around 1400 bound exoplanets with masses
exceeding $\sim 0.1$ Earth masses, including approximately 200 planets with masses less than 3 
Earth masses \citep{Penny2019}. Additionally, Roman will detect roughly 250 free-floating planets, 
with masses as low as that of Mars, including about 60 planets with masses below Earth mass
\citep{Johnson2020}.

In gravitational microlensing light curves, planets appear as a range of distinctive signal patterns.
These planetary signals are generated when a source star passes through the perturbation region
created by the caustic, which is induced by the gravitational influence of a planet. The size and
shape of the caustic vary depending on the separation between the planet and its host star, as
well as the mass ratio of the two. As a result, the planetary signals can take on a variety of
forms, depending on the planet's position and mass in relation to its host star. Furthermore, the
pattern of the planetary signal further varies depending on the path the source star takes relative
to the caustic, adding more complexity to the observed light curve.

\begin{table*}[t]
\small
\caption{Event coordinates and correspondence.  \label{table:one}}
\begin{tabular}{lccllcc}
\hline\hline
\multicolumn{1}{c}{Event}                      &
\multicolumn{1}{c}{(RA, DEC)$_{\rm J2000}$}    &
\multicolumn{1}{c}{$(l, b)$}                   &
\multicolumn{1}{c}{Other reference}            \\
\hline
 MOA-2022-BLG-091   &   (17:55:46.13, -32:20:29.18)  & $(-1^\circ\hskip-2pt .8118, -3^\circ\hskip-2pt .616)$   &   KMT-2022-BLG-0114  \\
 KMT-2024-BLG-1209  &   (17:26:27.11, -28:07:14.02)  & $(-1^\circ\hskip-2pt .5803,  3^\circ\hskip-2pt.9868)$   &   OGLE-2024-BLG-0777  \\
\hline
\end{tabular}
\end{table*}

As the discovery of microlensing planets increases, instances of planets exhibiting 
similar signal patterns are being identified. For example, \citet{Han2025} identified 
planets by analyzing three lensing events MOA-2022-BLG-033, KMT-2023-BLG-0119, and 
KMT-2023-BLG-1896 that displayed common short-term dip features near the peaks of highly 
magnified lensing light curves. Another group of planetary events with similar signal 
characteristics, reported by \citet{Han2024a}, includes KMT-2020-BLG-0757Lb, 
KMT-2022-BLG-0732Lb, KMT-2022-BLG-1787Lb, and KMT-2022-BLG-1852Lb. In these cases, the 
planetary signals appeared as extended negative deviations on the wings of the lensing 
light curves. Additionally, \citet{Han2024b} presented analyses on the light curves of 
two planetary events KMT-2021-BLG-2609 and KMT-2022-BLG-0303, for which the planetary 
signals appeared as as positive deviations on the wings of lensing light curves. Planetary 
events with similar traits were reported by \citet{Jung2021} for the lensing events 
OGLE-2018-BLG-0567 and OGLE-2018-BLG-0962 and by \citet{Ryu2024} for events 
OGLE-2017-BLG-1777 and OGLE-2017-BLG-0543. Studying planetary lensing events that show 
similar signal patterns is not only important for understanding the mechanisms behind 
these signals but also essential for identifying similar planetary signals in future 
discoveries.

In this paper, we analyze two microlensing events, MOA-2022-BLG-091 and KMT-2024-BLG-1209, 
which display anomalies with closely similar patterns.  We investigate the origins of these 
anomalies and the degeneracies that complicate their interpretation. The structure of this 
paper is as follows.  In Sect.~\ref{sec:two}, we outline the data used in our analyses, 
including details of the observations, the data reduction process, and the instrument 
specifications. Sect.~\ref{sec:three} provides a brief overview of the fundamentals of 
planetary microlensing and describes the modeling procedure employed. The results for each 
event are presented in detail in the subsequent sections: Sect.~\ref{sec:four} focuses on 
MOA-2022-BLG-091, while Sect.~\ref{sec:five} discusses KMT-2024-BLG-1209.  In Sect.~\ref{sec:six}, 
we identify the source stars for each event and estimate the angular radius of the Einstein 
ring.  In Sect.~\ref{sec:seven}, we describe the Bayesian analyses conducted to derive the 
physical parameters of the lenses and present these parameters. Finally, in Sect.~\ref{sec:eight}, 
we summarize the findings and conclude our study.

\section{Observations and data} \label{sec:two}

The planetary signals in the two lensing events were initially identified through an 
inspection of the light curves of KMTNet events KMT-2022-BLG-0114 and KMT-2024-BLG-1209. 
A subsequent review of data from other microlensing surveys revealed that KMT-2022-BLG-0114 
was also observed by the MOA survey, where it is designated as MOA-2022-BLG-091, while 
KMT-2024-BLG-1209 was additionally observed by the OGLE survey and designated as 
OGLE-2024-BLG-0777. Moreover, follow-up observations using two telescopes from the Las 
Cumbres Observatory (LCO) global telescope network covered the peak region of 
KMT-2024-BLG-1209. Our analysis is based on the combined datasets from these observations.

The event KMT-2022-BLG-0114 was announced by the KMTNet survey on March 23, 2022, while
the MOA survey issued an alert earlier, on March 14, 2022. For KMT-2024-BLG-1209, the
KMTNet survey issued an alert on March 31, 2024, followed by the OGLE survey on June 17,
2024. When an event is observed by multiple surveys, it is customary to adopt the event ID
designated by the group that first discovered it. Therefore, we designate the events as
MOA-2022-BLG-091 and KMT-2024-BLG-1209 throughout this work. Table~\ref{table:one} provides the
equatorial and Galactic coordinates of the events along with their corresponding survey
designations.

The three optical lensing surveys were conducted using multiple telescopes distributed across 
the Southern Hemisphere. The KMTNet survey operates three identical telescopes, located at the 
Siding Spring Observatory in Australia (KMTA), the Cerro Tololo Inter-American Observatory in 
Chile (KMTC), and the South African Astronomical Observatory in South Africa (KMTS). Each 
telescope has a 1.6-meter aperture and is equipped with a camera that provides a 4-square-degree 
field of view. The OGLE survey uses a 1.3-meter telescope at Las Campanas Observatory in Chile, 
equipped with a camera that provides a 1.4-square-degree field of view. The MOA survey operates 
a 1.8-meter telescope at Mt. John University Observatory in New Zealand, which is fitted with 
a camera covering a 2.2-square-degree field of view.  The LCO data covering the peak region of 
KMT-2024-BLG-1209 were taken using the 1.0~m telescopes from the Siding Spring Observatory 
in Australia (LCOA) and the Cerro Tololo Inter-American Observatory in Chile (LCOC).  Images 
from these followup observations were taken in the $I$ band.

Images from the KMTNet and OGLE surveys were primarily taken in the $I$ band, while those from 
the MOA survey were captured in the customized MOA-$R$ band, which covers a wavelength range 
of 609–1109 nm. A subset of images from all three surveys was taken in the $V$ band to measure 
the source colors of the events.  Image reduction and photometry for the microlensing events 
were conducted using the pipelines developed by the respective survey groups: the KMTNet data 
were processed with the pipeline from \citet{Albrow2009}, the OGLE data with the pipeline from 
\citet{Udalski2003}, and the MOA data with the pipeline described by \citet{Bond2001}. For the 
KMTNet and LCO data sets, we performed additional photometry using the code developed by 
\citet{Yang2024} to ensure optimal data quality.  For each dataset, we adjusted the error bars 
to achieve two objectives: first, they were consistent with the data scatter, and second, the 
value of $\chi^2$ per degree of freedom was equal to unity.  This error bar normalization 
followed the procedure outlined in \citet{Yee2012}.

\begin{table*}[t]
\caption{Lensing parameters of four degenerate solutions of MOA-2022-BLG-091.  \label{table:two}}
\begin{tabular}{lllllll}
\hline\hline
\multicolumn{1}{c}{Parameter}  &
\multicolumn{1}{c}{Sol A}      &
\multicolumn{1}{c}{Sol B}      &
\multicolumn{1}{c}{Sol C}      &
\multicolumn{1}{c}{Sol D}      \\
\hline
 $\chi^2$              &  $1727.8             $    &   $1726.7              $   &  $1733.4             $   &  $1731.7            $   \\
 $t_0$ (HJD$^\prime$)  &  $9657.729 \pm 0.012 $    &   $9657.785 \pm 0.017  $   &  $9657.699 \pm 0.0127$   &  $9657.706 \pm 0.008$   \\
 $u_0$ ($10^{-2}$)     &  $2.565 \pm 0.080    $    &   $2.614 \pm 0.087     $   &  $2.642 \pm 0.084    $   &  $2.590 \pm 0.84    $   \\
 $\te$ (days)          &  $34.68 \pm 0.86     $    &   $34.51 \pm 0.89      $   &  $34.53 \pm 0.87     $   &  $34.54 \pm 0.85    $   \\
 $s$                   &  $1.0029 \pm 0.0025  $    &   $1.0033 \pm 0.0023   $   &  $1.0132 \pm 0.0006  $   &  $1.0195 \pm 0.0027 $   \\
 $q$ ($10^{-3}$)       &  $4.55 \pm 0.48      $    &   $4.36 \pm 0.28       $   &  $3.38 \pm 0.31      $   &  $4.10 \pm 0.55     $   \\
 $\alpha$ (rad)        &  $4.632 \pm 0.018    $    &   $4.354 \pm 0.073     $   &  $4.811 \pm 0.034    $   &  $4.745 \pm0.004    $   \\
 $\rho$ ($10^{-3}$)    &  $ < 1.5             $    &   $< 1.5               $   &  $< 1.5              $   &  $< 1.5             $   \\
\hline
\end{tabular}
\tablefoot{ ${\rm HJD}^\prime = {\rm HJD}- 2450000$.  }
\end{table*}

\section{Fundamentals of planetary lensing and modeling} \label{sec:three}

A planet-induced caustic take different shape and appears at different locations depending 
on the projected separation from the host to the planet and the planet/star mass ratio
\citep{Gaudi2012}.  When the planet/star mass ratio is very small and the planet lies 
away from the Einstein ring, the planet forms two sets of caustics. One tiny caustic 
is located near the host star (central caustic), and the other is positioned at a position 
of ${\bf u}_{\rm c} = {\bf s} - 1/{\bf s}$ from the star (peripheral caustic).  Here, 
${\bf s}$ denotes the position vector of the planet relative to its host, with its length 
scaled to the angular Einstein radius $\thetae$.  When $s>1$ (wide planet), $u_{\rm c}>0$, 
and a peripheral caustic is located in the region between the star and the planet. On the 
other hand, when $s<1$ (close planet), $u_{\rm c}<0$, and peripheral caustics are located 
on the opposite side of the planet relative to the star. A wide planet creates a single 
four-cusp peripheral caustic, while a close planet creates two three-cusp caustics. For 
further details on the variation of caustics with respect to planetary separation and 
mass ratio, refer to \citet{Han2006} for peripheral caustics and \citet{Chung2005} for 
central caustics.

For giant planets with mass ratios on the order of $10^{-3}$ and located near the Einstein 
ring of their host stars, the central and planetary caustics merge into a single, large 
"resonant" caustic. This caustic forms a closed curve consisting of six concave segments 
that meet at six cusps. The cusps aligned along the planet-host axis are strong, producing 
significant deviations when the source passes through this region. In contrast, the cusps 
oriented perpendicular to the axis are weaker, resulting in relatively small perturbations 
when the source crosses these areas.

Modeling of the light curves was performed to determine a lensing solution, representing 
the set of lensing parameters that best account for the observed anomalies. For a lensing 
event in which the lens is composed of two masses and the source is a single star (2L1S 
event), the lensing light curve is described by seven basic parameters. The first three 
parameters $(t_0, u_0, \te)$ describe the lens-source approach. Specifically, $t_0$ 
represents the time of closest approach, $u_0$ denotes the lens-source separation at that 
moment (scaled to $\thetae$), and $\te$ is the event timescale, defined as the time required 
for the source to traverse $\thetae$. Two additional parameters, $(s, q)$, represent the 
projected separation and mass ratio between the binary lens components, respectively, and 
define the binary lens. Another parameter, $\alpha$, denotes the source incidence angle, 
which is the angle between the direction of the source's motion and the binary lens axis. 
Finally, the normalized source radius, $\rho$, is required to characterize the finite 
lensing magnifications during the source's crossing or close approach to the caustic.

Modeling was performed using a combination of grid and downhill methods.  For the lensing 
parameters $(s,q)$, where the magnification changes discontinuously with variations in 
these parameters, we used a grid approach with multiple initial values of $\alpha$.  The 
initial parameter ranges explored in the grid search were set to $-1.0 \leq \log s \leq 
1.0$ and $-5.0 \leq \log q \leq 1.0$, and were subsequently refined in a stepwise manner.  
For the source trajectory angle, 25 initial values uniformly spaced over the range from $0$ 
to $2\pi$.  For the remaining parameters, where the magnification changes smoothly, we applied 
a downhill method.  In the downhill approach, the Markov Chain Monte Carlo (MCMC) method was 
applied. In the first round of modeling, we constructed a $\chi^2$ map on the $s$--$q$--$\alpha$ 
parameter planes to examine potential degenerate solutions. We then refined the individual 
local solutions by allowing all parameters to vary. To assess the severity of the degeneracy, 
we compared the $\chi^2$ values of the fits for the local solutions. If the degeneracy was 
significant, we presented all local solutions and investigated the underlying causes of the 
degeneracies.  In the following sections, we present the result of analyses for the individual 
events.

\begin{figure}[t]
\includegraphics[width=\columnwidth]{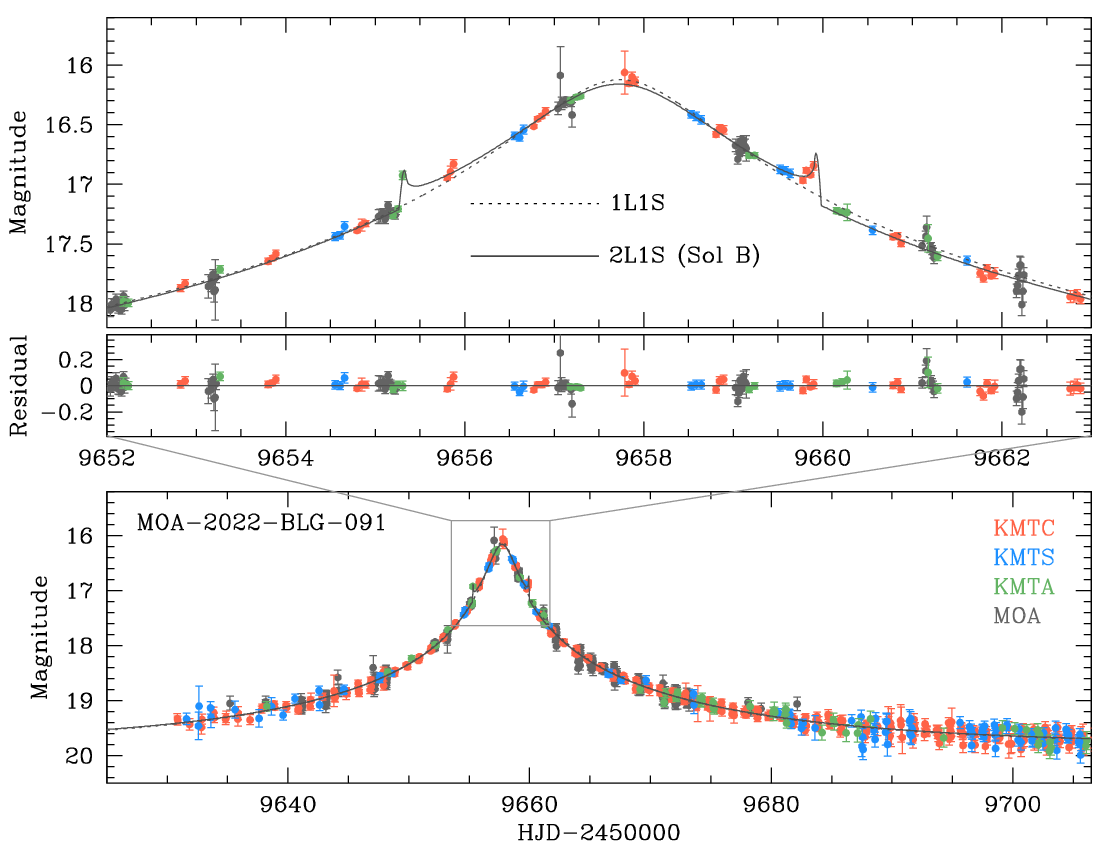}
\caption{
Light curve of lensing event MOA-2022-BLG-091. The lower panel displays the full light 
curve, while the upper panels provide a zoomed-in view of the region around the peak. The 
dotted and solid curves overlaid on the data points represent the 1L1S model and one of the 
2L1S models (solution B), respectively. The middle panel shows the residuals from the 2L1S 
solution. The colors of the data points are selected to correspond with the legends.
}
\label{fig:one}
\end{figure}

\begin{figure}[t]
\includegraphics[width=\columnwidth]{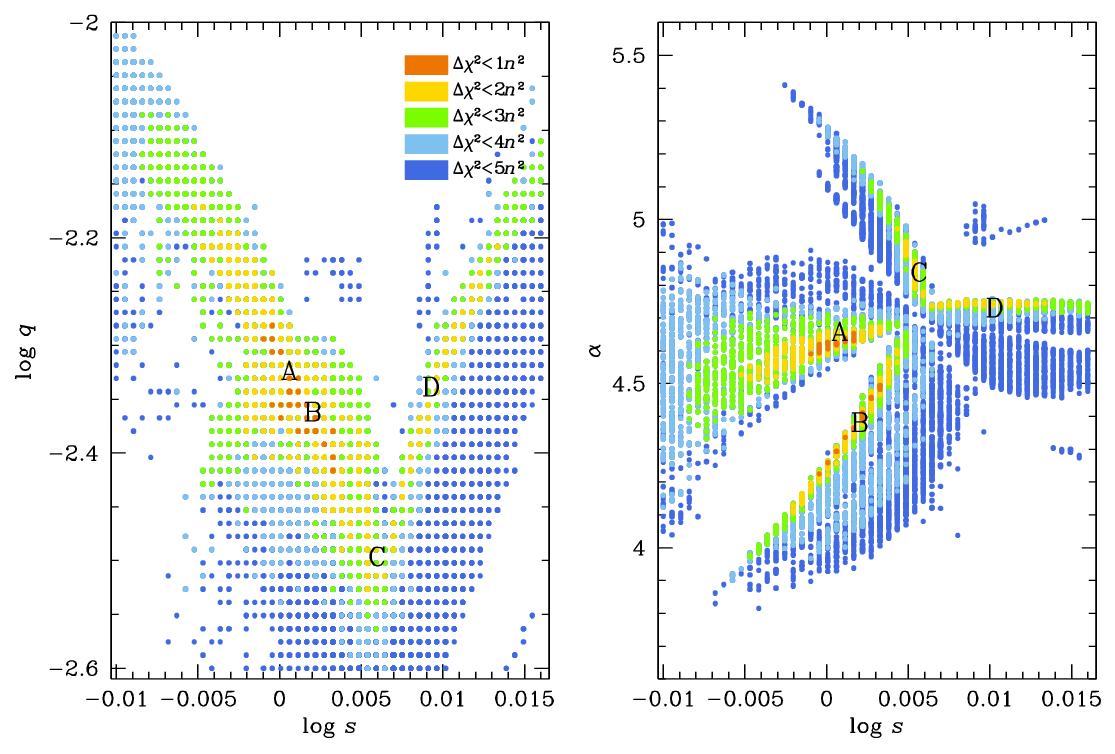}
\caption{
Locations of four degenerate solutions of MOA-2022-BLG-091 in the $\Delta\chi^2$ maps on the 
$\log s$–$\log q$ (left panel) and $\log s$–$\alpha$ (right panel) parameter planes. The color
coding represents regions with $\Delta\chi^2 < 1n^2$ (red), $\Delta\chi^2 < 2n^2$ (yellow), 
$\Delta\chi^2 < 3n^2$ (green), $\Delta\chi^2 < 4n^2$ (cyan), and $\Delta\chi^2 < 5n^2$ (blue), 
where $n=2$. 
}
\label{fig:two}
\end{figure}

\section{MOA-2022-BLG-091 } \label{sec:four}

Figure~\ref{fig:one} shows the lensing light curve of the event MOA-2022-BLG-091.  The event 
reached a moderately high magnification of $A_{\rm max}\sim 31$ at the peak.  At first glance, 
the light curve appears to be that of a single-lens single-source (1L1S) event with a smooth 
and symmetric shape.  However, upon closer inspection, we observed a weak deviation in the 
light curve.  The upper panel presents a close-up view of the peak region, illustrating that 
the light curve deviates from the 1L1S model (dotted curve), which was derived by excluding 
the data points near the anomaly.

The most noticeable deviations in the light curve occurred in two distinct regions: around 
${\rm HJD} \sim 2459655.3$, approximately 2.4 days before the peak, and around ${\rm HJD} 
\sim 2459659.8$, about 2.1 days after the peak. The time interval between these two anomalies 
is roughly 4.5 days. The first deviation was recorded by a single KMTA data point, while the 
second was captured by four KMTC data points.  In both cases, the lensing magnifications exhibit 
abrupt variations, suggesting that the anomalies may have been caused by the source's caustic 
crossings.  However, unlike typical caustic crossings, where deformations occur both when the 
source crosses caustics and when it is inside the caustic, for example, the planet-induced 
anomaly in the light curve of the lensing event KMT-2021-BLG-1150 \citep{Han2023}, there is 
little deviation in the region between the caustic spikes.  This results in anomaly characteristics 
that differ from the typical case.

\begin{figure}[t]
\includegraphics[width=\columnwidth]{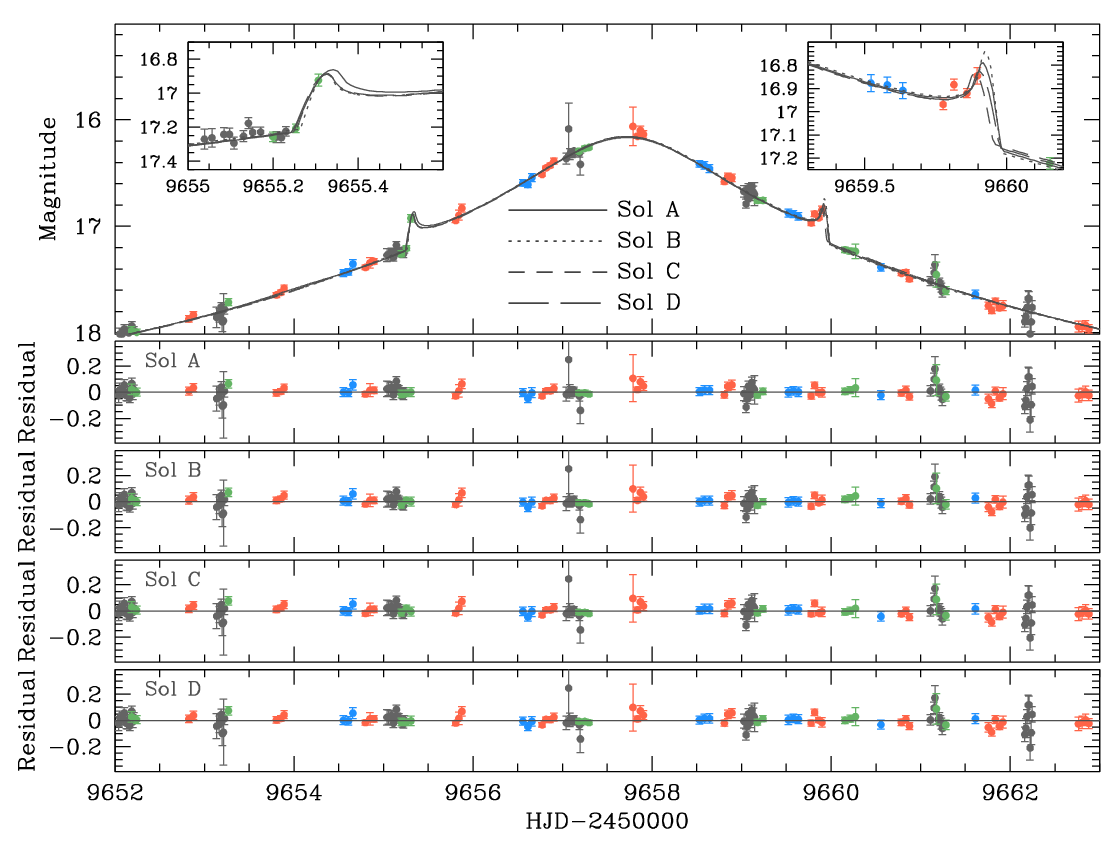}
\caption{
Model curves of MOA-2022-BLG-091 in the region around the anomaly.  The insets in the top 
panel provide blowup of the regions around the anomalies induced by the source's caustic 
crossings.  The lower panels present the residuals corresponding to the four solutions.
}
\label{fig:three}
\end{figure}

Modeling of the light curve indicated that the anomaly was produced by a planetary companion 
to the lens, although the interpretation is complicated by the existence of multiple possible 
solutions.  We identified four degenerate solutions, which were labeled as “Sol A,” “Sol B,” 
“Sol C,” and “Sol D.” Figure~\ref{fig:two} shows the locations of these solutions in the 
$\Delta\chi^2$ maps on the $\log s$–$\log q$ (left panel) and $\log s$–$\alpha$ (right panel) 
parameter planes.  Among these, Sol B provides the best fit, but the $\chi^2$ differences among 
the solutions are less than 6.7, indicating a strong degeneracy among them. Figure~\ref{fig:three} 
displays the model curves of the four solutions in the region of the anomaly, illustrating that 
all solutions reproduce the anomaly with nearly equal accuracy.

The complete sets of lensing parameters for each solution, along with the corresponding 
$\chi^2$ values of the fits, are provided in Table~\ref{table:two}. Among these solutions, 
Sol C has a mass ratio of $q \sim 3.4 \times 10^{-3}$, which is smaller than the mass ratios 
of the other solutions, which range from $q \sim 4.1 \times 10^{-3}$ to $q \sim 4.6 \times 
10^{-3}$. Although the mass ratios show slight variations, all are several times greater 
than the Jupiter/Sun mass ratio, indicating that the companion to the lens is a planetary 
mass object.  The planet-host separations for all solutions are close to $s = 1$, implying 
that the planet is located near the Einstein ring of the primary lens. The event timescales, 
$\te \sim 35$ days, are consistent across the solutions. One key distinguishing feature of 
the solutions is the source incidence angle, which ranges from 3.4 to 4.6 radians. Due to 
limited coverage of the caustic crossings, the exact value of the normalized source radius 
could not be determined, and only an upper limit of $\rho_{\text{max}} \sim 1.5 \times 
10^{-3}$ can be established.

\begin{figure}[t]
\includegraphics[width=\columnwidth]{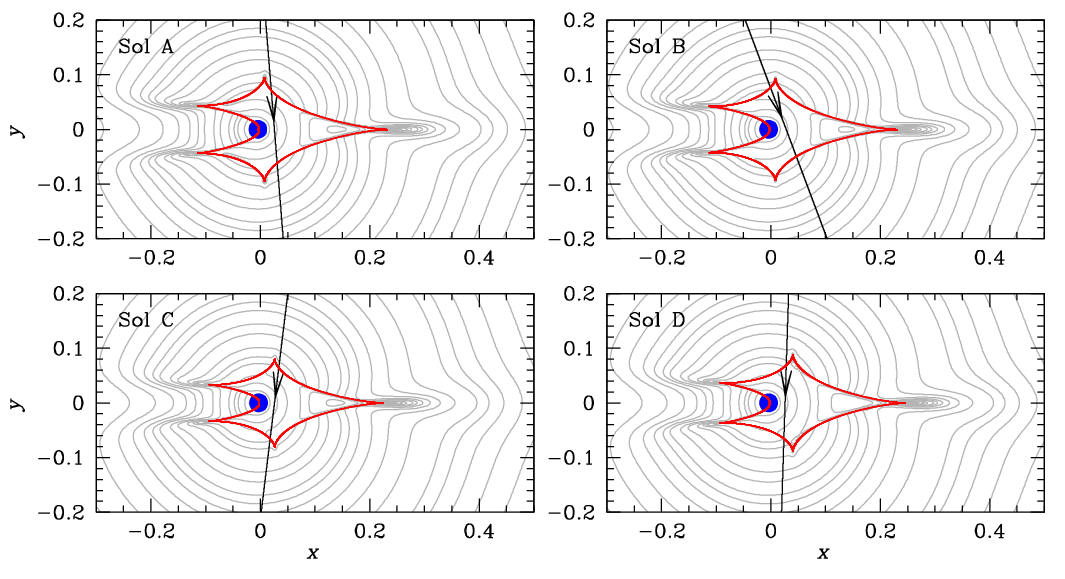}
\caption{
Configurations of lens system for the four degenerate solutions of MOA-2022-BLG-091. In 
each panel, the cuspy closed curve represents the caustic, while the arrowed line denotes 
the source trajectory. The gray contours surrounding the caustic represent equi-magnification 
contours.  The blue filled circle presents the location of the planet host. 
}
\label{fig:four}
\end{figure}

To understand the cause of the degeneracies among the solutions, we examined the lens 
system configurations. Figure~\ref{fig:four} shows these configurations, where the red 
cusped shape represents the caustic, the blue filled circle marks the position of the 
planet host, and the arrowed line indicates the source trajectory. All solutions form a 
single resonant caustic, which appears similar due to the comparable values of $s$ and 
$q$ across the solutions. The primary differences arise from the source trajectories.  
In Sol~A, the source crosses to the right of both the upper and lower cusps of the caustic, 
while in Sol~D, it crosses to the left of both cusps. For Sol~B, the source trajectory 
passes to the left of the upper cusp and the right of the lower cusp, while for Sol~C, the 
source passes to the right of the upper cusp and the left of the lower cusp.  This degeneracy 
represents a previously unrecognized type that has not been observed in earlier interpretations 
of planetary signals.  Therefore, it is important to consider this degeneracy when analyzing 
similar planetary perturbations in future events.

Since the similarities among the model light curves result from chance rather than intrinsic 
similarities in the caustic structures of specific lens configurations, the degeneracies 
could have been lifted under more favorable observational conditions. In this event, the 
KMTNet survey operated with a 1.0-hour cadence, in contrast to the 15-minute cadence of 
the Roman RGES survey. Given the high photometric precision of space-based observations, 
such degeneracies are expected to be resolved in the RGES survey.

\begin{figure}[t]
\includegraphics[width=\columnwidth]{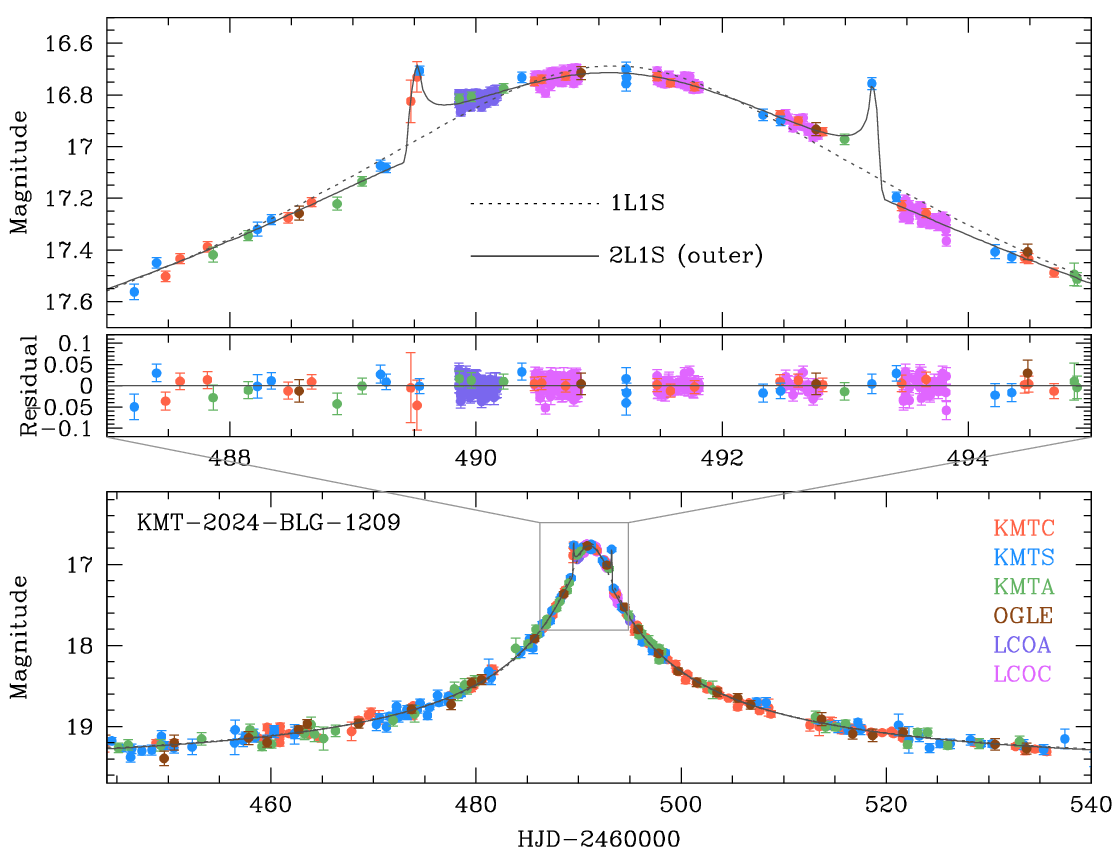}
\caption{
Light curve of the lensing event KMT-2024-BLG-1209.  Notations are same as those in 
Fig.~\ref{fig:one}.
}
\label{fig:five}
\end{figure}

\section{KMT-2024-BLG-1209} \label{sec:five}

The lensing light curve of KMT-2024-BLG-1209 is shown in Figure~\ref{fig:five}. A closer 
inspection of the peak region reveals an anomaly characterized by two caustic spikes: the 
first occurred before the peak at ${\rm HJD} \sim 2460489.5$, and the second occurred after 
the peak at ${\rm HJD} \sim 2460492.2$, with an approximate time gap of 2.7 days between them. 
The event reached a moderately high magnification of $A_{\rm max} \sim 45$ at its peak. While 
the anomaly in KMT-2024-BLG-1209 is more pronounced than the subtle anomaly in MOA-2022-BLG-091, 
the two events share several similarities.  First, both events exhibit weak caustic-crossing 
signatures, occurring before and after the peak, respectively.  Second, both events reach 
moderately high magnifications at their peaks, and the region between the caustic spikes 
shows little deviation.

\begin{figure}[t]
\includegraphics[width=\columnwidth]{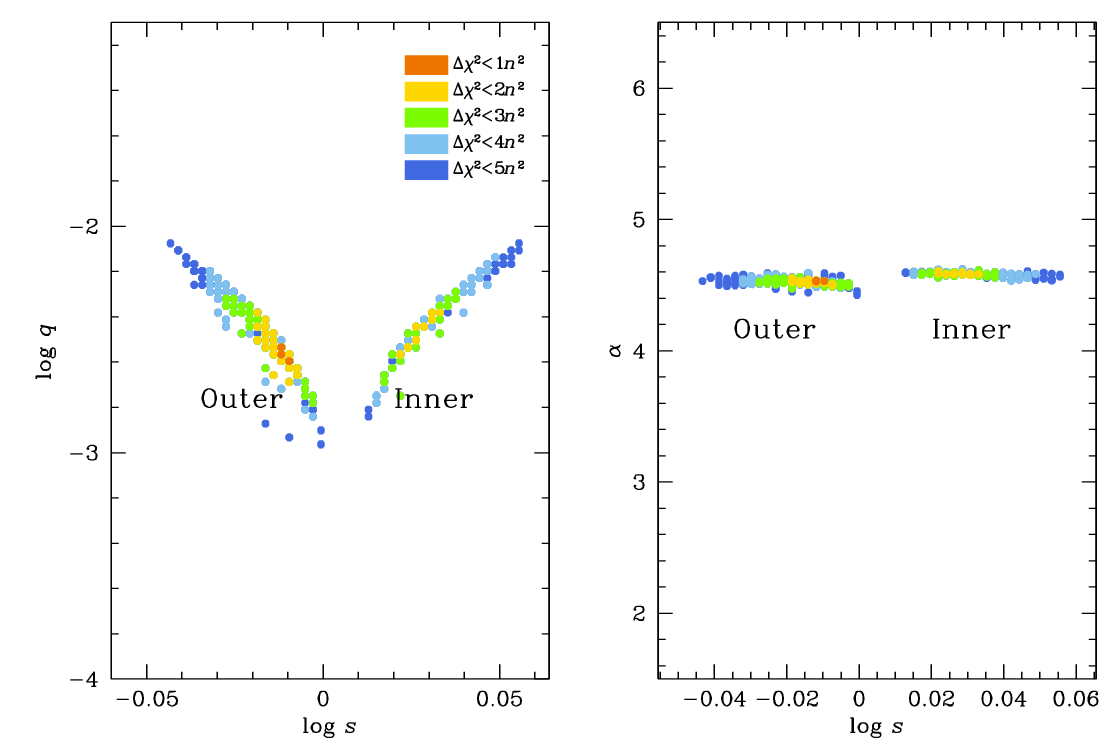}
\caption{
$\Delta\chi^2$ maps on the $\log s$–$\log q$ (left panel) and $\log s$–$\alpha$
for KMT-2024-BLG-1209. Notations are same as those in Fig.~\ref{fig:two}.
}
\label{fig:six}
\end{figure}

Given the similarities in the anomaly characteristics, we model the light curve of 
KMT-2024-BLG-1209 using a 2L1S configuration. We identified two local solutions, with the 
degeneracy between them stemming from a different origin compared to MOA-2022-BLG-091. 
Figure~\ref{fig:six} shows the locations of the two local solutions in the $\Delta\chi^2$ 
maps on the parameter space. These solutions are designated as "inner" and "outer," with 
the reasoning for these labels provided below.

The complete sets of lensing parameters for both solutions, along with the corresponding 
$\chi^2$ values, are presented in Table~\ref{table:three}. For the inner solution, the 
binary lens parameters are $(s, q)_{\rm in} \sim (1.06, 3.33 \times 10^{-3})$, while for 
the outer solution, the parameters are $(s, q)_{\rm out} \sim (0.97, 2.90 \times 10^{-3})$.  
The degeneracy is significant, with $\Delta\chi^2 = 2.0$.  The mass ratio, which is slightly 
more than three times that of Jupiter to the Sun, suggests that the lens companion lies in 
the planetary mass regime.  Given the relatively long timescale of the event, we examined 
the possibility of detecting microlens parallax effects caused by Earth's orbital motion 
around the Sun \citep{Gould1992a}. However, securely measuring the microlens parallax was 
challenging, primarily due to significant photometric uncertainties at low magnifications.

\begin{figure}[t]
\includegraphics[width=\columnwidth]{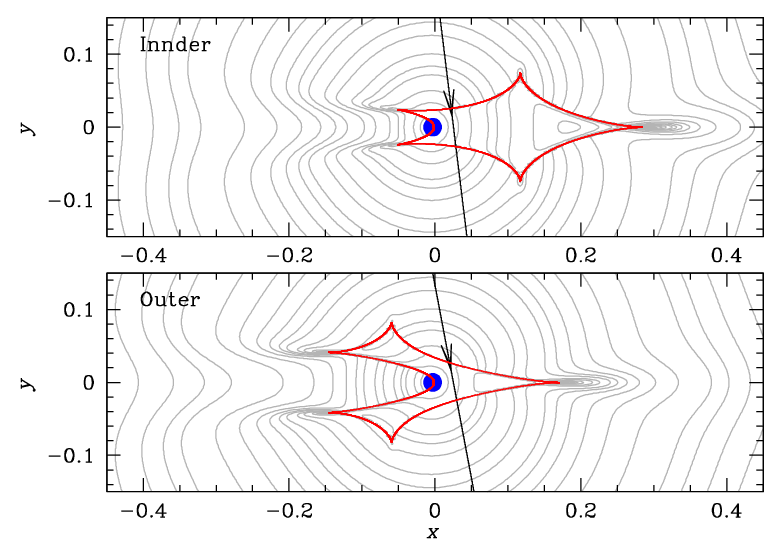}
\caption{
Lens system configurations for the two degenerate solutions of KMT-2024-BLG-1209. 
Notations are same as those in Fig.~\ref{fig:four}.
}
\label{fig:seven}
\end{figure}

To understand the cause of the degeneracy between the two solutions, we compared their 
lens system configurations, as depicted in Figure~\ref{fig:seven}. Both solutions produce 
a single resonant caustic, but the shapes differ. In the inner solution, the caustic 
appears to result from the merger of the planetary caustic on the planet side, generated 
by a wide planet, with the central caustic. In contrast, the outer solution’s caustic 
seems to form from the combination of two planetary caustics on the opposite side of the 
planet, created by a close planet, along with the central caustic. In the inner solution, 
the source crosses the inner region of the planetary caustic relative to the planet host, 
while in the outer solution, the source passes through the outer region of the planetary 
caustic. Based on this distinction, we have labeled the solutions as the "inner" and 
"outer" solutions \citep{Gaudi1997}. The normalized source radius, $0.8 \times 10^{-3}$, 
was determined from the relatively well-covered caustic region, and the measured event 
time scale is approximately $\te \sim 72$~days.

\begin{table}[t]
\caption{Lensing parameters of two solutions of KMT-2024-BLG-1209.\label{table:three}}
\begin{tabular*}{\columnwidth}{@{\extracolsep{\fill}}lllll}
\hline\hline
\multicolumn{1}{c}{Parameter}        &
\multicolumn{1}{c}{Inner}            &
\multicolumn{1}{c}{Outer}            \\
\hline
 $\chi^2$               &  $1731.3           $   &  $1729.3           $  \\  
 $t_0$ (HJD$^\prime$)   &  $491.126 \pm 0.009$   &  $491.141 \pm 0.007$  \\
 $u_0$ (10$^{-2}$)      &  $2.57 \pm 0.14    $   &  $2.49 \pm 0.14    $  \\
 $\te$ (days)           &  $70.11 \pm 3.77   $   &  $73.13 \pm 3.81   $  \\
 $s$                    &  $1.0599 \pm 0.0036$   &  $0.9724\pm 0.0034 $  \\
 $q$  (10$^{-3}$)       &  $3.33 \pm 0.29    $   &  $2.90 \pm 0.24    $  \\
 $\alpha$ (rad)         &  $4.5957 \pm 0.0066$   &  $4.5294 \pm 0.0083$  \\
 $\rho$ (10$^{-3}$)     &  $0.75 \pm 0.20    $   &  $0.73 \pm 0.14    $  \\
\hline             
\end{tabular*}
\tablefoot{ ${\rm HJD}^\prime = {\rm HJD}- 2460000$.  }
\end{table}

The similarity between the two model curves arises due to the inner-outer degeneracy. 
This is evidenced by the fact that the planet separations of the inner ($s_{\rm in}$) 
and outer ($s_{\rm out}$) solutions satisfy the following relation \citep{Hwang2022, 
Gould2022}: 
\begin{equation} 
\langle s \rangle = s^\dagger; \ \  
\langle s \rangle = \sqrt{s_{\rm in} \times s_{\rm out}}, \ \ 
s^\dagger = \sqrt{u_{\rm anom}^2 + 4} \pm u_{\rm anom}.
\label{eq1} 
\end{equation} 
Here, $\langle s \rangle$ represents the geometric mean of $s_{\rm in}$ and $s_{\rm out}$, 
$u_{\rm anom}^2 = \tau^2 + u_0^2$, $\tau_{\rm anom} = (t-t_{\rm anom})/\te$, and $t_{\rm anom}$ 
denotes the time of the anomaly.  The sign in the last term is "$+$" for a positive anomaly 
and "$-$" for a negative anomaly.  For KMT-2024-BLG-1209, the planet induced a positive 
deviation, and therefore the sign is "$+$".  
For the values $(s_{\rm in}, s_{\rm out}) = (1.0596, 0.9699)$, 
we calculate the geometric mean $\langle s \rangle = 1.014$. Using the parameters $(t_0, u_0, 
\te, t_{\rm anom}) \sim (491.142, 0.0248, 72.363, 491.0)$, we obtain $s^\dagger = 1.013$. 
The close agreement between the values of $\langle s \rangle$ and $s^\dagger$ confirms that 
the similarity between the model curves of the two solutions is a result of the inner-outer 
degeneracy.

\section{Source stars and angular Einstein radii} \label{sec:six}

In this section, we define the source stars for the events. Identifying the source is crucial 
not only for fully characterizing the event but also for determining the angular Einstein 
radius. The angular Einstein radius is related to the angular source radius, $\theta_*$, 
through the relation
\begin{equation}
\thetae = {\theta_* \over \rho}, 
\label{eq2}
\end{equation}
where $\theta_*$ is inferred from the source type, and the normalized source radius, $\rho$, 
is measured through modeling.

\begin{figure}[t]
\includegraphics[width=\columnwidth]{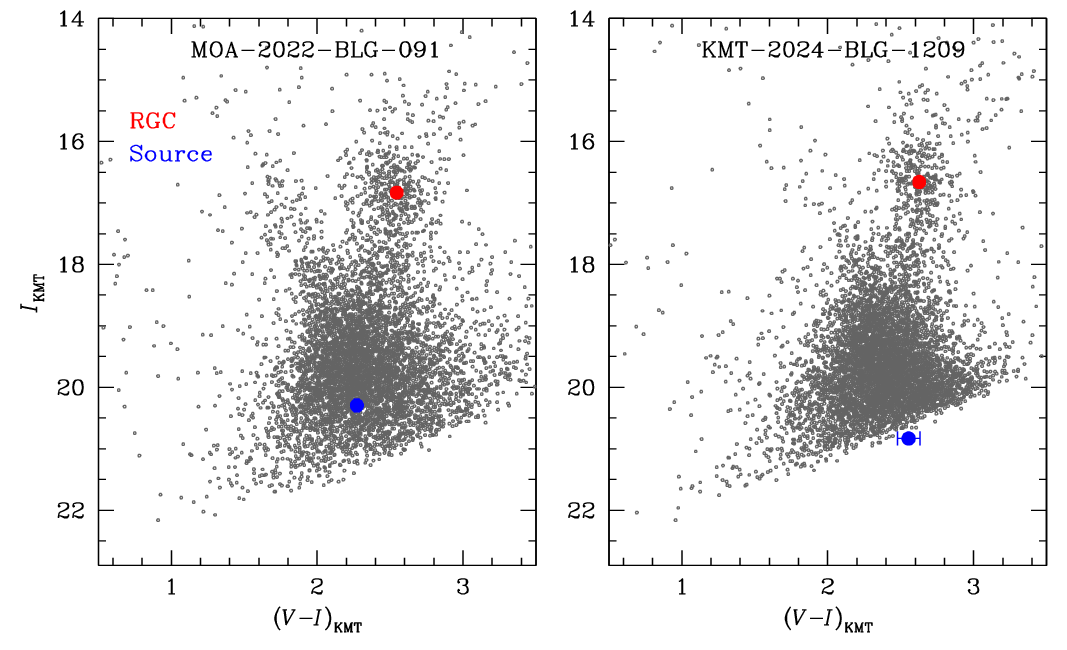}
\caption{
Locations of source stars (blue filled dots) in the instrumental color-magnitude diagrams 
for nearby stars around the source stars of MOA-2022-BLG-091 (left panel) and KMT-2024-BLG-1209 
(right panel). Also marked are the centroids of red giant clump (RGC, red filled dots) used 
for magnitude calibration. 
}
\label{fig:eight}
\end{figure}

To determine the source type, we first estimated the source color, $(V - I)$. As an initial 
step, we measured the instrumental $V$- and $I$-band magnitudes by fitting the light curves 
in the respective passbands to the lensing model. For this measurement, we utilized photometric 
data processed with the pyDIA code \citep{Albrow2017}. Figure~\ref{fig:eight} presents the 
positions of the source stars in the instrumental color-magnitude diagrams (CMDs) for nearby 
stars around the sources of MOA-2022-BLG-091 (left panel) and KMT-2024-BLG-1209 (right panel). 
The CMDs were constructed using the pyDIA code, which was also employed to measure the source 
magnitudes.  To minimize the spatial variation of extinction, the CMDs were constructed for 
stars lying within 1.7 arcmin from the source.  Table~\ref{table:four} provides the measured 
instrumental color and magnitude of the source star, $(V - I, I)_{\rm S}$.  The uncertainties 
of the source color and magnitude are estimated by considering the uncertainty of the RGC 
position, $\sigma(V-I, I)_{\rm RGC}=(0.04, 0.02)$ \citep{Bensby2013, Nataf2013}.

In the second step, we calibrated the source color and magnitude using the red giant clump 
(RGC) centroid in the CMD as a reference \citep{Yoo2004}. The RGC centroid was chosen as 
the reference because its intrinsic (dereddened) color and magnitude, $(V - I, I)_{{\rm RGC},0}$, 
have been well established in previous studies \citep{Bensby2013, Nataf2013}.  The dereddened 
source color and magnitude, $(V - I, I)_{\rm S,0}$, were then determined by applying the 
measured offsets in color and magnitude, $\Delta (V - I, I)$, between the source and the RGC 
centroid in the CMD, following the relation
\begin{equation} 
(V - I, I)_{{\rm S},0} = (V - I, I)_{{\rm RGC},0} + \Delta(V - I, I).
\label{eq3} 
\end{equation}
Table~\ref{table:four} presents the estimated values of $(V - I)_{\rm S,0}$ along with 
$(V - I, I)_{\rm RGC}$ and $(V - I, I)_{{\rm RGC},0}$. The results indicate that the source 
of MOA-2022-BLG-091 is a late G-type main-sequence star, while the source of KMT-2024-BLG-1209 
is an early K-type main-sequence star.

\begin{table}[t]
\caption{Source parameters and angular Einstein radii.  \label{table:four}}
\begin{tabular*}{\columnwidth}{@{\extracolsep{\fill}}lllll}
\hline\hline
\multicolumn{1}{c}{Parameter}             &
\multicolumn{1}{c}{MOA-2022-BLG-091}      &
\multicolumn{1}{c}{KMT-2024-BLG-1209}     \\
\hline
 $(V-I)_{\rm S}$             &   $2.272 \pm 0.015 $    &  $2.554 \pm 0.078  $  \\
 $I_{\rm S}$                 &   $20.295 \pm 0.002$    &  $20.833 \pm 0.003 $  \\
 $(V-I, I)_{\rm RGC}$        &   $(2.545, 16.833) $    &  $(2.627, 16.660)  $  \\
 $I_{{\rm RGC},0}$           &     14.339              &    14.533             \\
 $(V-I)_{{\rm S},0}$         &   $0.787 \pm 0.043 $    &  $0.987 \pm 0.087  $  \\
 $I_{S,0}$                   &   $17.801 \pm 0.020$    &  $18.705 \pm 0.020 $  \\
  Type                       &    G8V                  &   K3V                 \\
 $\theta_*$ ($\mu$as)        &   $0.945 \pm 0.077 $    &  $0.783 \pm 0.089  $  \\
 $\thetae$ (mas)             &   $> 0.63          $    &  $1.133 \pm 0.378  $  \\
 $\mu$ (mas/yr)              &   $> 6.63          $    &  $5.63 \pm 1.88    $  \\
\hline
\end{tabular*}
\tablefoot{$(V-I)_{{\rm RGC},0}=1.016$.   }
\end{table}

Using the measured color and magnitude, we estimated the angular source radius. To achieve 
this, we first converted the $V-I$ color into the $V-K$ color using the color-color relation 
from \citet{Bessell1988}. We then derived the angular source radius, $\theta_*$, from the 
\citet{Kervella2004} relation between $(V-K, I)$ and $\theta_*$. With the estimated angular 
source radius, we proceeded to estimate the angular Einstein radius, $\thetae$, using the 
relation in Eq.~(\ref{eq2}).  Additionally, we calculated the relative lens-source proper 
motion using the relation $\mu = \thetae/\te$. In Table~\ref{table:four}, we present the 
estimated values of $\theta_*$, $\thetae$, and $\mu$ for the events. For event MOA-2022-BLG-091, 
where only the upper limit of the normalized source radius was constrained, we provide the lower 
limits for $\thetae$ and $\mu$.

\begin{figure}[t]
\includegraphics[width=\columnwidth]{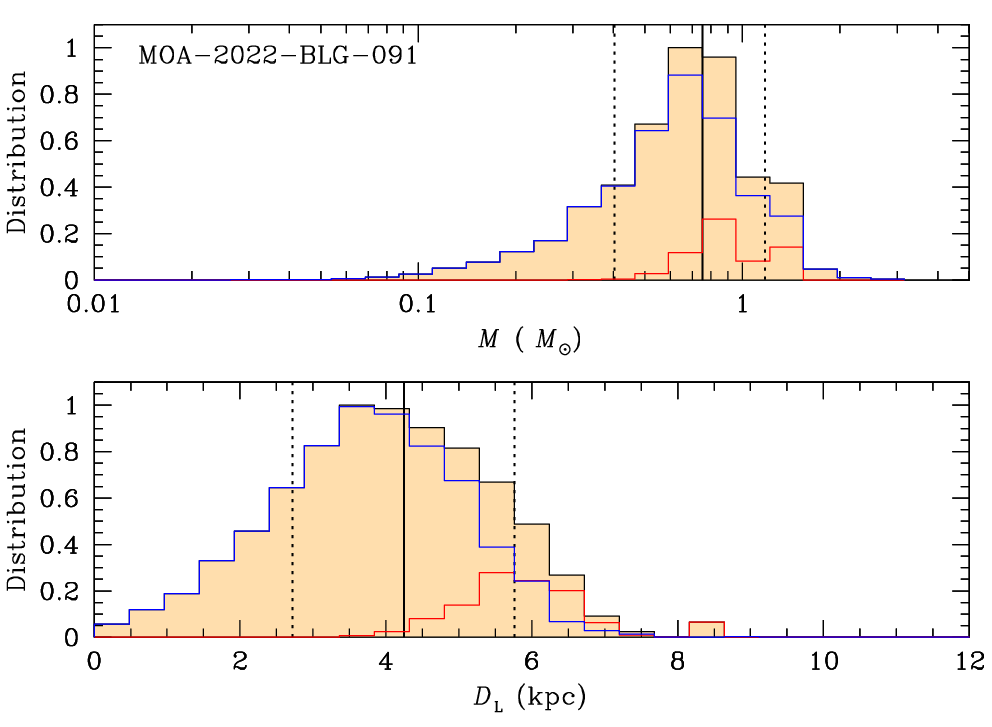}
\caption{
Bayesian posteriors of the lens mass (upper panel) and distance (lower panel) for
MOA-2022-BLG-091. In each panel, the solid vertical line represents the median of the posterior
distribution, while the two dotted lines mark the 16th and 84th percentiles of the distribution. 
The contributions of the disk and bulge lens populations are shown separately as blue and 
red curves, respectively.
}
\label{fig:nine}
\end{figure}

\section{Physical lens parameters} \label{sec:seven}

We estimated the physical parameters of the lens mass ($M$) and distance ($\dl$) through 
a Bayesian analysis. This analysis incorporated constraints from the lensing observables, 
combined with priors from the lens mass function and the Galactic model.  For the events 
under consideration, the lensing observables that provide information on the physical lens 
parameters include the event time and the angular Einstein radius.  These quantities are 
related to the physical parameters through the expression: 
\begin{equation}
\te = {\thetae \over \mu};\qquad
\thetae = \sqrt{\kappa M \pi_{\rm rel}}. 
\label{eq4}
\end{equation}
Here $\kappa=4G/(c^2 {\rm AU})$, $\pi_{\rm rel} = \pi_{\rm L} - \pi_{\rm S} = {\rm AU}(D_{\rm
L}^{-1} - D_{\rm S}^{-1})$, with $\ds$ representing the distance to the source.

The Bayesian analysis begins by generating a large number of synthetic events through a Monte
Carlo simulation. For each simulated event, the physical parameters $(M_i, D_{{\rm L},i}, \mu)$
were assigned based on the lens mass function and the Galactic model, which define the physical
and dynamical distributions of the lens and source. In this simulation, we adopted the model mass
function of \citet{Jung2021} and the Galaxy model of \citet{Jung2022}. Using the assigned physical 
parameters, we then calculated the corresponding lens observables based on the relations in
Eq.~(\ref{eq4}).  Using the simulated events, a Bayesian posterior was derived by constructing 
the distributions of $M$ and $\dl$, with each simulated event assigned a weight. The assigned 
weight was calculated as 
\begin{equation}
w_i = \exp\left( -{\chi_i^2\over 2}\right); \qquad 
\chi_i^2 = \left[ {t_{{\rm E},i} - \te \over \sigma(\te) }\right]^2 +
\left[ {\theta_{{\rm E},i} - \thetae \over \sigma(\thetae) }\right]^2.
\label{eq5}
\end{equation}
Here, $(\te, \thetae)$ denote the measured lensing observables, while $[\sigma(\te),
\sigma(\thetae)]$ represent their respective measurement uncertainties. For MOA-2022-BLG-091,
for which only a lower limit on the angular Einstein radius, $\theta_{{\rm E},\text{min}}$, is
available, we impose a constraint of $\theta_{{\rm E},i} > \theta_{{\rm E},\text{min}}$.

\begin{figure}[t]
\includegraphics[width=\columnwidth]{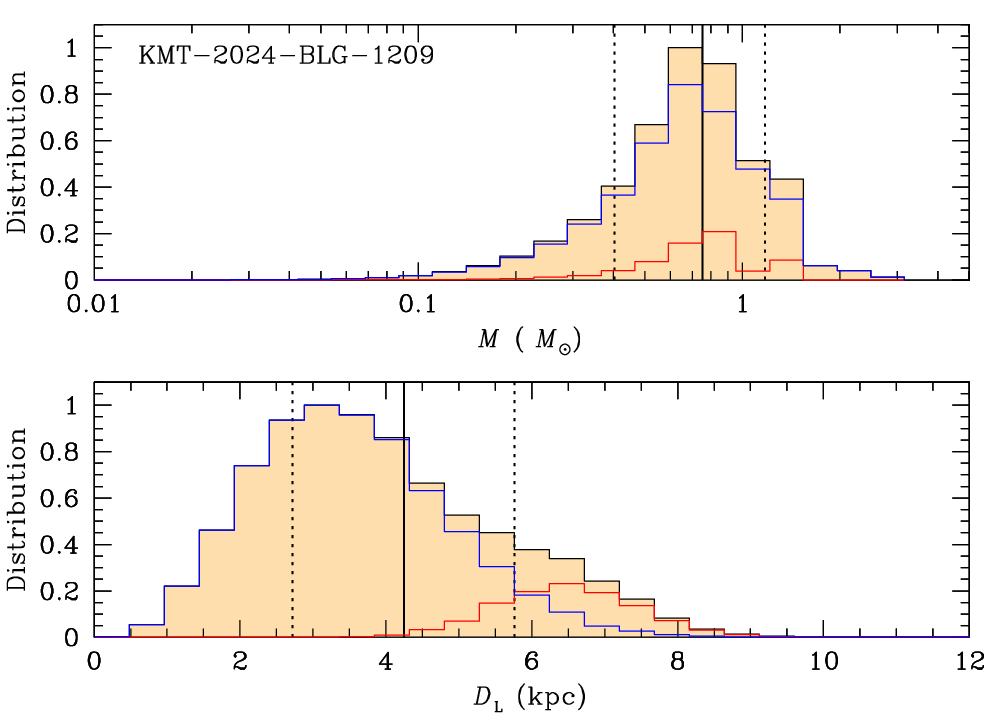}
\caption{
Posterior of the lens mass and distance for KMT-2024-BLG-1209. Notations are same as those
in Fig.~\ref{fig:nine}. 
}
\label{fig:ten}
\end{figure}

\begin{table}[t]
\caption{Physical lens parameters.\label{table:five}}
\begin{tabular*}{\columnwidth}{@{\extracolsep{\fill}}lllll}
\hline\hline
\multicolumn{1}{c}{Parameter}      &
\multicolumn{1}{c}{MOA-2022-BLG-091}      &
\multicolumn{1}{c}{KMT-2024-BLG-1209}    \\
\hline
  $M_{\rm h}$ ($M_\odot$)     &  $0.75^{+0.42}_{-0.35}$                &  $0.77^{+0.46}_{-0.34}$              \\  [0.6ex]
  $M_{\rm p}$ ($M_{\rm J}$)   &  $3.59^{+2.00}_{-1.67}$                &  $2.64^{+1.57}_{-1.16}$              \\  [0.6ex]
  $\dl$ (kpc)                 &  $4.25^{+1.52}_{-1.53}$                &  $3.93^{+2.03}_{-1.41}$              \\  [0.6ex]
  $a_\perp$ (AU)              &  $3.63^{+1.30}_{-1.31}$ \ \ (Sol A)    &  $3.92^{+2.03}_{-1.41}$ \ \ (inner)  \\  [0.6ex]
                              &  $3.44^{+1.92}_{-1.60}$ \ \ (Sol B)    &  $2.51^{+1.50}_{-1.10}$ \ \ (outer)  \\  [0.6ex]
                              &  $2.67^{+1.45}_{-1.24}$ \ \ (Sol C)    &  --                                  \\  [0.6ex]
                              &  $3.26^{+1.80}_{-1.51}$ \ \ (Sol D)    &  --                                  \\  [0.6ex]
  $p_{\rm disk}$              &  $87\%                $                &  $86\%                $              \\  [0.6ex]
  $p_{\rm bulge}$             &  $13\%                $                &  $14\%                $              \\  [0.6ex]

\hline             
\end{tabular*}
\end{table}

In Figures~\ref{fig:nine} and \ref{fig:ten}, we present the posterior distributions of the 
lens mass and distance for the events MOA-2022-BLG-091 and KMT-2024-BLG-1209, respectively. 
In Table~\ref{table:five}, we provide the estimated values for the host and planet masses 
($M_{\rm h}$ and $M_{\rm p}$), the distance to the system, and the projected separation of 
the planet from the host ($a_\perp$).  We adopt the median of the posterior distribution as 
the representative value and define the uncertainty as the range between the 16th and 84th 
percentiles of the distribution.  For each event, we present the values of $a_\perp$ 
corresponding to the degenerate solutions, which result in different projected separations. 
The table also includes the probabilities for the planetary system being located in the 
Galactic disk ($p_{\rm disk}$) or bulge ($p_{\rm bulge}$).

The physical parameters of the two events exhibit similar characteristics in several aspects.  
First, the planetary companions are giant planets with masses two to four times that of Jupiter 
in our solar system. Second, the hosts of both planets are main-sequence stars of an early K-type 
spectral type. Third, the planet is located beyond the host's snow line, which is given by 
$a_{\rm snow} \sim 2.7~{\rm AU} (M/M_\odot) \sim 2.0~{\rm AU}$. Fourth, both planetary systems 
are very likely to be in the Galactic disk.

\section{Summary and conclusion} \label{sec:eight}

We analyzed the microlensing events KMT-2022-BLG-0114 and KMT-2024-BLG-1209, whose light
curves exhibited anomalies with very similar characteristics. These anomalies, occurring near the
peaks of the lensing light curves with moderately high magnifications, are characterized by weak
caustic-crossing features with minimal deformation while the source was inside the caustic.

Through detailed modeling of the light curves, we found that the anomalies in both lensing events
indicate a planetary origin. These anomalies arose from the source crossing the resonant caustic
formed by a planet, with an incidence angle close to a right angle. However, precise interpretation
of the anomalies was hindered by various degeneracies. For KMT-2022-BLG-0114, the challenge
stemmed from a previously unrecognized degeneracy associated with the uncertainty in the
incidence angle of the source’s trajectory relative to the planet-host axis. In the case of
KMT-2024-BLG-1209, the analysis was complicated by the previously known degeneracy, which
results in ambiguity between solutions where the source crosses either the inner or outer side of
the caustic relative to the planet host.

The physical parameters of the two planetary systems, estimated through a Bayesian analysis,
exhibit similar characteristics in several aspects. In both events, the companions to the lenses are
giant planets with masses ranging from two to four times that of Jupiter, and their hosts are early
K-type main-sequence stars. Each planet is located beyond the snow line of its host star, and both
planetary systems are highly likely to reside in the Galactic disk.

The degeneracy in KMT-2024-BLG-1209 is difficult to resolve due to intrinsic similarities in 
the caustic structures between the inner and outer solutions. In contrast, the newly identified 
degeneracy in MOA-2022-BLG-091 arises by chance rather than from inherent features of the caustic 
itself. Consequently, this degeneracy is expected to be resolved by the upcoming Roman RGES 
microlensing survey, which will offer higher observational cadence and improved photometric 
precision from space.

\begin{acknowledgements}
This research has made use of the KMTNet system operated by the Korea Astronomy and Space Science 
Institute (KASI) at three host sites of CTIO in Chile, SAAO in South Africa, and SSO in Australia. 
Data transfer from the host site to KASI was supported by the Korea Research Environment Open 
NETwork (KREONET).  C.Han acknowledge support from KASI under the R\&D program (project No. 
2024-1-832-01) supervised by the Ministry of Science and ICT.  The MOA project is supported by 
JSPS KAKENHI Grant Number JP24253004, JP26247023,JP16H06287 and JP22H00153.  J.C.Y., I.G.S., and 
S.J.C. acknowledge support from NSF Grant No. AST-2108414.  C.R. was supported by the Research 
fellowship of the Alexander von Humboldt Foundation.  W.Zang acknowledges the support from the 
Harvard-Smithsonian Center for Astrophysics through the CfA Fellowship.  The LCO research uses 
data obtained through the Telescope Access Program (TAP), which has been funded by the TAP member 
institutes. This work makes use of observations from the Las Cumbres Observatory global telescope 
network H.Y., W.Zang, Y.T., S.M. and W.Zhu acknowledge support by the National Natural Science 
Foundation of China (Grant No. 12133005).  
\end{acknowledgements}


\begin{thebibliography}{}
\bibitem[Albrow et al.(2009)]{Albrow2009} Albrow, M., Horne, K., Bramich, D. M., et al. 2009, \mnras, 397, 2099
\bibitem[Albrow et al.(2017)]{Albrow2017} Albrow, M. 2017, MichaelDAlbrow/pyDIA: Initial Release on Github,Versionv1.0.0, Zenodo, doi:10.5281/zenodo.268049
\bibitem[Bensby et al.(2013)]{Bensby2013}  Bensby, T. Yee, J.C., Feltzing, S. et al. 2013, \aap, 549, A147
\bibitem[Bessell \& Brett(1988)]{Bessell1988} Bessell, M. S., \& Brett, J. M. 1988, \pasp, 100, 1134
\bibitem[Bond et al.(2001)]{Bond2001}  Bond, I. A., Abe, F., Dodd, R. J., et al. 2001, \mnras, 327, 868
\bibitem[Chung et al.(2005)]{Chung2005} Chung, S.-J., Han, C., Park, B.-G., et al. 2005, \apj, 630, 535     
\bibitem[Gaudi(2012)]{Gaudi2012} Gaudi, B. S, 2012, \araa, 50m 411
\bibitem[Gaudi \& Gould(1997)]{Gaudi1997} Gaudi, B. S., \& Gould, A. 1997, \apj, 486, 85
\bibitem[Gould(1992)]{Gould1992a} Gould, A. 1992, \apj, 392, 442
\bibitem[Gould \& Loeb(1992)]{Gould1992b} Gould, A., \& Loeb, A. 1992, \apj, 396, 104
\bibitem[Gould et al.(2022)]{Gould2022} Gould, A., Han, C., Zang, W., et al. 2022, \aap, 664, A13
\bibitem[Han(2006)]{Han2006} Han, C. 2006, \apj, 638, 1080
\bibitem[Han et al.(2020)]{Han2020} Han, C., Udalsk, A., Gould, A., et al. 2020, \aj, 159, 91
\bibitem[Han et al.(2023)]{Han2023} Han, C., Jung, Y. K., Bond, I. A. 2023, \aap, 675, A36 
\bibitem[Han et al.(2024b)]{Han2024a} Han, C., Bond, I. A., Lee, C.-U., et al. 2024q, \aap, 687, A225 
\bibitem[Han et al.(2024b)]{Han2024b} Han, C., Albrow, M. D., Lee, C.-U., et al. 2024b, \aap, 689, A209
\bibitem[Han et al.(2025)]{Han2025} Han, C., Bond, I. A., Jung, Y. K., et al. 2025, \aap, in press, arXiv:2501.0
\bibitem[Hwang et al.(2022)]{Hwang2022} Hwang, K.-H., Zang, W., Gould, A., et al. 2022, \aj, 163, 43
\bibitem[Johnson et al.(2020)]{Johnson2020} Johnson, S. A., Penny, M., Gaudi, B. S., et al. 2020, \aj, 160, 123
\bibitem[Jung et al.(2021)]{Jung2021} Jung, Y. K., Han, C., Udalski, A., et al. 2021, \aj, 161, 293
\bibitem[Jung et al.(2022)]{Jung2022} Jung, Y. K., Zang, W., Han, C., et al. 2022, \aj, 164, 262 
\bibitem[Jung et al.(2024)]{Jung2024} Jung, Y. K., Hwang, K.-H., Yang, H., et al. 2024, \aj, 168, 152
\bibitem[Kervella et al.(2004)]{Kervella2004} Kervella, P., Th\'evenin, F., Di Folco, E., \& S\'egransan, D. 2004, \aap, 426, 29
\bibitem[Kim et al.(2021)]{Kim2021} Kim, H.-W., Hwang, K.-H., Gould, A., et al. 2021, \aj, 162, 15
\bibitem[Kim et al.(2016)]{Kim2016} Kim, S.-L., Lee, C.-U., Park, B.-G., et al. 2016, JKAS, 49, 37
\bibitem[Kondo et al.(2023)]{Kondo2023} Kondo, I., Sumi, T., Koshimoto, N., et al. 2023, \aj, 165, 54
\bibitem[Mao \& Paczy\'nski(1991)]{Mao1991}  Mao, S., \& Paczy\'nski, B. 1991, \apj, 374, 37
\bibitem[Mr\'oz et al.(2018)]{Mroz2018} Mr\'oz, P., Ryu, Y.-H., Skowron, J., et al. 2018, \aj, 155, 121 
\bibitem[Mr\'oz et al.(2019)]{Mroz2019} Mr\'oz, P., Udalski, A., Bennett, D. P., et al. 2019, \aap, 622, A201
\bibitem[Mr\'oz et al.(2020)]{Mroz2020} Mr\'oz, P., Poleski, R., Han, C., et al. 2020, \aj, 159, 262
\bibitem[Nataf et al.(2013)]{Nataf2013} Nataf, D. M., Gould, A., Fouqu\'e, P. et al. 2013, \apj, 769, 88
\bibitem[Penny et al.(2019)]{Penny2019} Penny, M. T., Gaudi, B. S., Kerins, E., et al. 2019, \apjs, 241, 3
\bibitem[Poleski et al.(2020)]{Poleski2020} Poleski, R., Suzuki, D., Udalski, A., et al. 2020, \aj, 159, 261
\bibitem[Ryu et al.(2024)]{Ryu2024} Ryu, Y.-H., Udalski, A., Yee, J. C., et al. 2024, \aj, 167, 88
\bibitem[Sumi et al.(2003)]{Sumi2003} Sumi, T., Abe, F., Bond, I. A., et al. 2003, \apj, 591, 204
\bibitem[Udalski(2003)]{Udalski2003} Udalski, A. 2003, Acta Astron., 53, 291
\bibitem[Udalski et al.(2015)]{Udalski2015} Udalski, A., Szyma\'nski, M. K., \& Szyma\'nski, G. 2015, Acta Astron., 65, 
\bibitem[Yang et al.(2024)]{Yang2024} Yang, H., Yee, J. C., Hwang, K.-H., et al. 2024, \mnras, 528, 11
\bibitem[Yee et al.(2012)]{Yee2012} Yee, J. C., Shvartzvald, Y., Gal-Yam, A., et al.\ 2012, \apj, 755, 102
\bibitem[Yee et al.(2021)]{Yee2021} Yee, J. C., Zang, W., Udalski, A., et al. 2021, \aj, 162, 180
\bibitem[Yoo et al.(2004)]{Yoo2004} Yoo, J., DePoy, D.L., Gal-Yam, A. et al. 2004, \apj, 603, 13
\bibitem[Zhai et al.(2024)]{Zhai2024} Zhai, R., Poleski, R., Zang, W., et al. 2024, \aj, 167, 162

\end{thebibliography}
\end{document}